\documentclass[smallextended]{svjour3}       
\smartqed  
\usepackage{graphicx}
\usepackage{amssymb,amsmath}

\renewcommand{\le}{\leqslant}

\newcommand{\kk}{\mathbf{k}}

\begin{document}

\title{Metal--Insulator Transition in the Hubbard Model: Correlations and Spiral Magnetic Structures
}

\titlerunning{Metal--Insulator Transition in the Hubbard Model...}        

\author{Marat~A.~Timirgazin \and
	Petr~A.~Igoshev    \and
        Anatoly~K.~Arzhnikov  \and
        Valentin~Yu.~Irkhin
}


\institute{M.A. Timirgazin, A.K. Arzhnikov \at
              Physical-Technical Institute, 426000, Kirov str. 132, Izhevsk, Russia \\
              \email{timirgazin@gmail.com}           
           \and
           P.A. Igoshev \at
              Institute of Metal Physics, Russian Academy of Sciences, 620990 Ekaterinburg, Russia \\
              Ural Federal University, 620002 Ekaterinburg, Russia
              \email{igoshev\_pa@imp.uran.ru}
	   \and
           V.Yu. Irkhin \at
              Institute of Metal Physics, Russian Academy of Sciences, 620990 Ekaterinburg, Russia\\       
          \email{valentin.irkhin@imp.uran.ru}
}

\date{Received: date / Accepted: date}

\maketitle

\begin{abstract}
The metal--insulator transition (MIT) for the square, simple cubic, and body-centered cubic lattices is investigated within the $t$---$t'$ Hubbard model 
at half--filling  
by using both the generalized for the case of spiral order Hartree--Fock approximation (HFA) and Kotliar--Ruckenstein slave-boson approach. It turns out that magnetic scenario of MIT becomes superior over non-magnetic one. The electron correlations lead to some suppression of the spiral phases in comparison with HFA. We found the presence of metallic antiferromagnetic (spiral) phase in the case of three-dimensional lattices. 


\end{abstract}

\section{Introduction}
\label{intro}
Metal--insulator transitions (MIT's) are intensively studied starting from the 1940s until now \cite{1990:Mott}. However, the quantitative description of the MIT's and sometimes the qualitative understanding of the physical phenomena determining these transitions in definite systems still remain unsatisfactory \cite{1998:Imada}.

MIT investigation within the Hubbard model clarifies that electron correlations cause a gap formation in the electron paramagnetic spectrum, which leads to the metal--insulator transition at half-filling at finite values of the Coulomb interaction constant $U$ \cite{1964:Hubbard} (Mott scenario). Later, the description of such a transition was improved by the use of the many-electron X-operator approach \cite{1975:Zaitsev,2004:Irkhin} and  within the dynamical mean-field theory (DMFT), which predicts a first-order MIT \cite{1994:Rozenberg,2003:Onoda}.

It is well known that a strong tendency to antiferromagnetism is characteristic for the half-filled band, so that the magnetic (Slater) scenario of MIT should be considered besides the Mott scenario. Experimentally the magnetic scenario of MIT is observed in V$_2$O$_{3-y}$ and NiS$_{2-x}$Se$_x$ \cite{1998:Imada}. The account of the nearest-neighbor electron hopping only (with the transfer integral $t$) for the bipartite lattices results in exponentially small dielectric gap of the Slater type (antiferromagnetic subbands) appearing at any weak interaction $U$ between electrons, manifesting the formation of insulating state. When the next-nearest-neighbor hopping is taken into account (with the transfer integral $t'$) the antiferromagnetic (AF) metal--insulator transition occurs at a finite interaction. In the case of small $t'$ the first-order MIT occurs with the transition point determined by $1/U_\textrm{c} = \rho\ln(t/t')$, where $\rho$ is the density of states on the Fermi level \cite{1984:Katsnelson}. The DMFT studies of the AF state confirm the first-order transition at small $t'$  but predict the second-order transition for large $t'$ \cite{1999:Chitra}. For very large $t'\sim t$ the AF state on the Bethe lattice was shown to be strongly suppressed so that the paramagnetic MIT emerges from the magnetic region~\cite{2009:Peters}.

Usually MIT is studied taking into account only the checkerboard AF ordering, being usually assumed to dominate near half-filling. It was shown in Refs. \cite{2010:Igoshev,2013:Igoshev} that the incommensurate magnetic structures in the Hubbard model are stabilized in the form of spin-spiral states in a wide concentration region. The Hartree--Fock investigation of MIT with account of spiral magnetic states was carried out in Ref. \cite{2012:Timirgazin} and there was shown that a region of spiral metallic phase is present between paramagnetic metal and AF insulator in the square lattice at large values of $t'$. Later it was shown that spiral magnetic states can be stabilized at half-filling in cubic lattices as well \cite{2015:Igoshev,2015:JPCM}. Thus an analysis of the role of incommensurate magnetic states in MIT should be performed for cubic lattices.

We apply the Kotliar--Ruckenstein slave-boson approach (SBA) \cite{1986:Kotliar} to take into account the many-electron nature of electronic states beyond HFA. In the saddle point approximation, this method is qualitatively close to the known Gutzwiller approximation. Therefore, the ground state energy obtained is in a good agreement with the quantum Monte Carlo and exact diagonalization calculations~\cite{1992:Fresard}. 

\section{Method and results}
\label{Results}
We consider the spin-spiral state within the Hubbard model considering the nearest and next-nearest neighbor hopping approximation for the one-electron spectrum $t_{\kk}$. 
We treat a correlation-induced MIT in a half-filled band (other kinds of MIT, e.g. in disordered systems, are beyond our consideration). 
To treat  the spiral magnetic order formation we apply the generalized Kotliar--Ruckenstein slave-boson approach  \cite{2013:Igoshev,2015:JPCM,1992:Fresard} for the case of half-filled band and directly minimize the ground state thermodynamical potential with respect to spiral wave vector $\bf Q$. The calculations are based on the fermionic part of the effective Hamiltonian $H_{\rm f}$ 
\begin{equation}
	\label{Hf}
	H_{\rm f} = \sum_{\kk\sigma\sigma'} \left(z_{\sigma}z_{\sigma'}t_{\sigma\sigma'}(\kk)+\lambda_\sigma\delta_{\sigma\sigma'}\right)c^+_{\kk\sigma}c_{\kk\sigma'},
\end{equation}
where $t_{\sigma\sigma'}(\kk) = (1/2)(t_{\kk+\mathbf{Q}/2}+t_{\kk-\mathbf{Q}/2})\delta_{\sigma\sigma'} + (1/2)(t_{\kk+\mathbf{Q}/2}-t_{\kk-\mathbf{Q}/2})\delta_{\sigma,-\sigma'}$
\begin{equation}
	\label{z_def2}
	z_\sigma=(1-d^2-p_\sigma^2)^{-1/2} (ep_\sigma+p_{\bar\sigma}d) (1-e^2-p_{\bar\sigma}^2)^{-1/2}
\end{equation}
is the correlation one-electron band narrowing and $\lambda_\sigma$ is the one-electron shift.   
In the Eq. (\ref{z_def2}) the average probability amplitudes in quantum many-electron states representation (empty $e$, singly occupied $p_\sigma$, doubly occupied $d$) are calculated in a mean-field manner.
In the insulating magnetically ordered state the chemical potential $\mu$ should be positioned in the gap of one-electron spectrum $E_{\pm}(\kk)$,  which is given by eigenvalues of $H_{\rm f}$ (see Eq.~(\ref{Hf}))
\begin{equation}
	\max_{\kk} E_{-}(\kk) < \mu < \min_{\kk}E_{+}(\kk).
\end{equation}
For $0\le t' \le t/2$  we run over values of $U$ starting from zero until metal--insulator transition is found.
In the case of second-order magnetic phase transition the determination of critical $U_c/t$ is not sufficiently precise, thus  we determine it by calculating the generalized magnetic susceptibility within the SBA following the result of Ref. \cite{1991:Li} (a generalized Overhauser criterion). 

We have constructed the magnetic phase diagrams for square, simple cubic and body-centered cubic lattices in terms of variables $U/t$ and $t'/t$. 

\subsection{Square lattice}
\label{sec:square}
An intriguing issue is the existence of metallic magnetically ordered phase between paramagnetic metal and magnetically ordered insulating phases:
the HFA based result is a sequence of second-order transitions paramagnetic metal --- AF metal --- AF insulator at large $t'=t/2$ being changed by the first order transition at moderate $t'=0.2t$ \cite{1996:Kondo} (this was confirmed within the quantum Monte--Carlo calculations \cite{1997:Duffy}). The obtained results were verified in more detail within HFA in Ref. \cite{2010:Yu}. It was found that the metallic AF state is unstable at moderate $0.08t\lesssim t'\lesssim0.38t$. The account of many-electron nature of electronic states within Kotliar--Ruckenstein slave-boson approximation indicates the instability of this state at not very small $t'\gtrsim0.14t$ \cite{2000:Yang}. It is also interesting that at large $t'$ not only the checkerboard AF state was found but the so-called collinear AF state as well, which corresponds to the spiral structure with $\mathbf{Q}=(0,\pi)$ \cite{2010:Yu,2006:Imada,2013:Yamada}.

\begin{figure}[h]
  \includegraphics[height=0.45\textwidth]{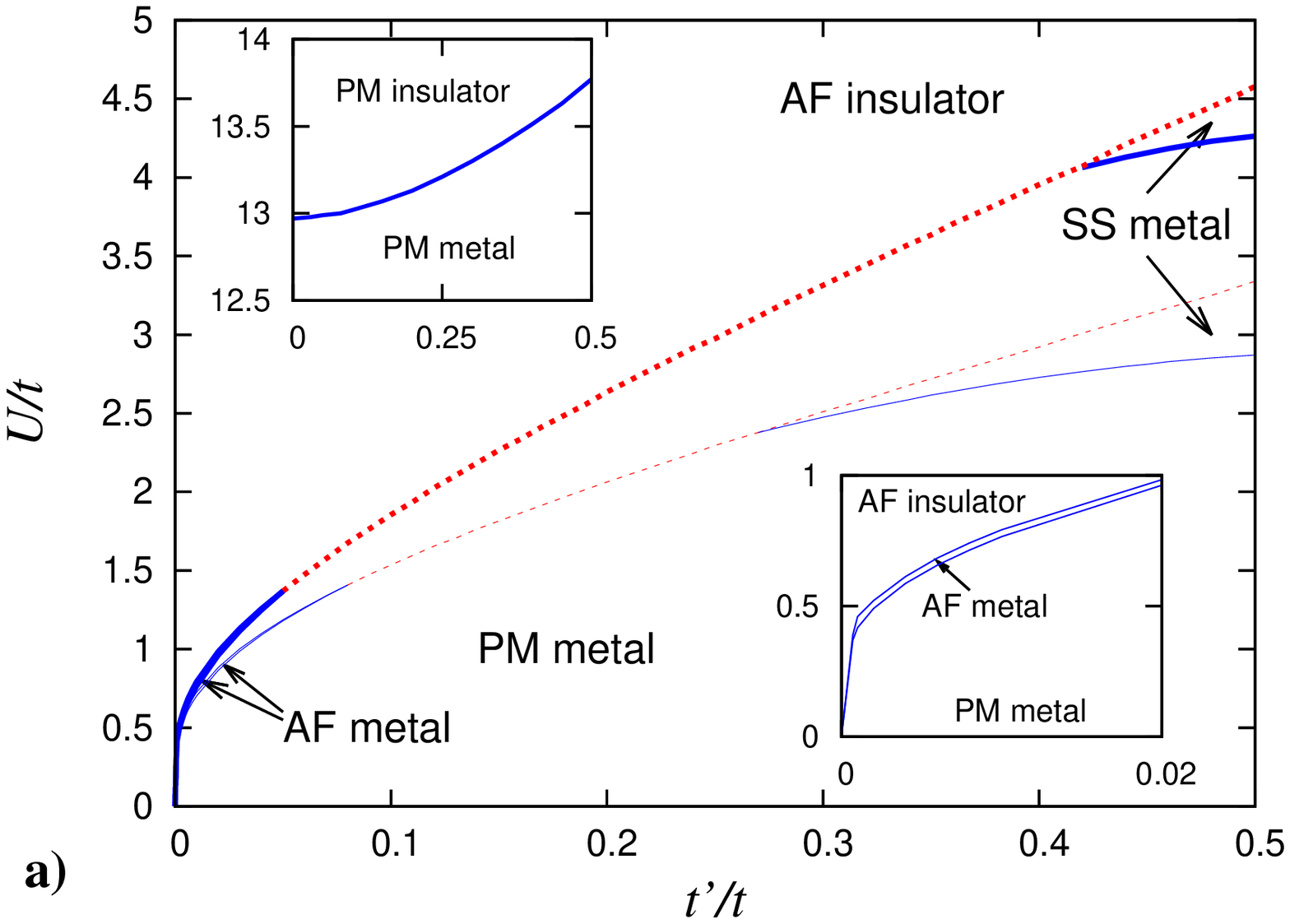}\includegraphics[height=0.45\textwidth]{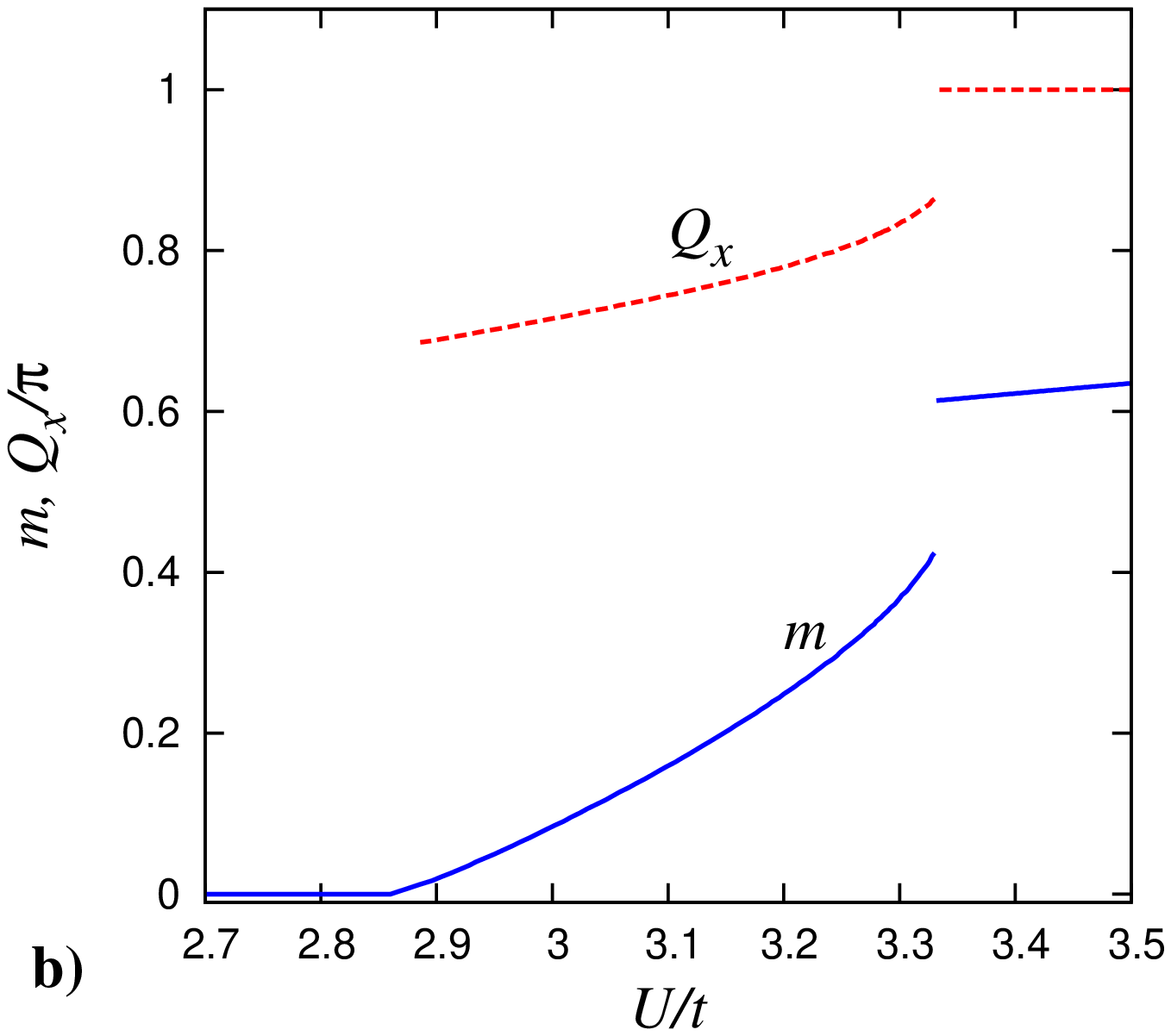}
  \caption{a) Ground state magnetic phase diagram of the Hubbard model at half-filling in $t'/t$---$U/t$ plane for the
square lattice. Thick (thin) lines present the result of SBA (HFA). Solid (dashed) lines present the second (first) order transition lines. The lower inset shows the same phase diagram obtained within SBA at small $t'$ more particularly. The upper inset shows the result of Mott scenario within the Brinkman--Rice criterion.
b) The dependence of magnetization $m$ (solid lines) and wave vector $Q_x$ (dashed lines) vs $U/t$ in HFA for $t'=0.5t$, $\mathbf{Q}=(Q_x,\pi)$. 
}\label{fig:square_phase_diagram}
\end{figure}

The resulting ground state phase diagram within both HFA and SBA for the square lattice is presented in Fig. \ref{fig:square_phase_diagram}a. Depending on a value of $t'$ the sequences of the metal--insulator and magnetic transitions have 3 scenarios.
(i) The increase of $U$ forces the system to go from paramagnetic metallic state into
AF metallic state through the second-order phase transition (see the lower inset to Fig. \ref{fig:square_phase_diagram}a). Further increase of $U$ results in the electronic spectrum gap opening, and the second-order phase transition into AF insulator state takes place ($t'\lesssim0.05t$ in SBA);
(ii) the paramagnetic -- antiferromagnetic transition and MIT coincide and are of the first order ($0.05t\lesssim t'\lesssim0.42t$);
(iii) the paramagnetic metallic state goes into the spin-spiral metallic state with $\mathbf{Q}=(\pi-\delta,\pi)$ (lattice constant is taken as unity) through the second-order phase transition ($t'\gtrsim0.42t$). 
Here and below $\delta$ generally depends on the values of model parameters ($U$, $t'/t$), so that only wave vector form is fixed. 
Further increase of $U$ forces the first-order metal--insulator transition from the spin-spiral metallic state into the AF insulator state.  
The change in position of the intervals within SBA in comparison with HFA reduces to a quantitative expansion of (ii) $t'$ region at the cost of the other regions constriction.
Fig. \ref{fig:square_phase_diagram}b illustrates the behavior of magnetization $m$ and wave vector $\mathbf{Q}$ depending on $U/t$ for $t'/t=0.5$ in HFA. One can see a jump of both values at first order transition between spiral metallic and antiferromagnetic insulator states. 

The first-order transitions in the (ii) and (iii) scenarios imply a possibility of magnetic phase separation which is common in vicinity of half-filling for the model considered \cite{2015:JPCM}. The determination of its $U$-boundaries is complicated since one should go beyond the half-filling in order to calculate the dependence of chemical potential on the electron concentration. Our study shows that the phase separation at half-filling, if present, cannot be detected within the accuracy of our calculations. 

The upper inset to Fig. \ref{fig:square_phase_diagram}a presents the paramagnetic metal--insulator transition obtained by the Brinkman--Rice (BR) criterion $U=U_{\rm BR}\equiv -16\sum_{\mathbf{k}}t_{\mathbf{k}}f(t_{\mathbf{k}})$, where $t_{\mathbf{k}}$ is a bare electronic spectrum \cite{1970:BR};  
we state that generally the Mott scenario is found to be irrelevant at the parameters studied, since characteristic values of $U_{\rm BR}$ are much larger than ones for magnetic metal--insulator transition.

Now we compare our results with the previous works. We see qualitative and quantitative coincidence in HFA results with Ref. \cite{2010:Yu}. The region of collinear antiferromagnetic state found in that work at large $t'$ is proved to be just a part of the spiral states region with variable $\mathbf{Q}$.
The Hartree--Fock results of Ref.~\cite{2000:Yang} are not confirmed. In contrast to this paper we do not find AF metal phase in all the range of $t'$. At the same time our SBA results agree qualitatively with Ref.~\cite{2000:Yang}.
However there are quantitative dissimilarities: we find AF metal phase only at $t'\lesssim0.05t$ while authors of Ref.~\cite{2000:Yang} found it also at $0.06t\leqslant t'<0.14t$ with the AF metal--insulator transition being of the first order.

\subsection{Simple cubic and bcc lattices}

\begin{figure}[h]
  \includegraphics[width=0.6\textwidth]{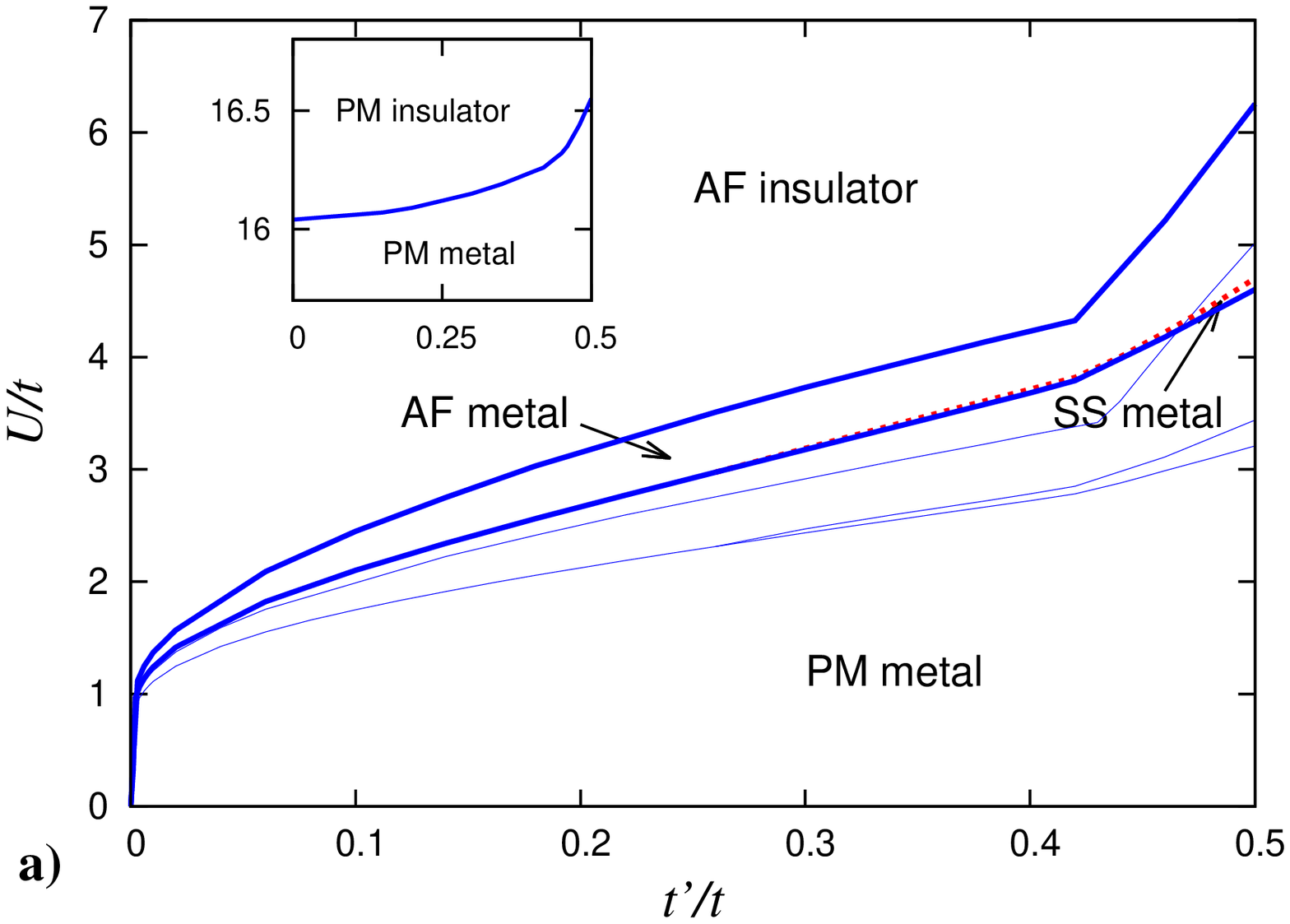}\includegraphics[width=0.6\textwidth]{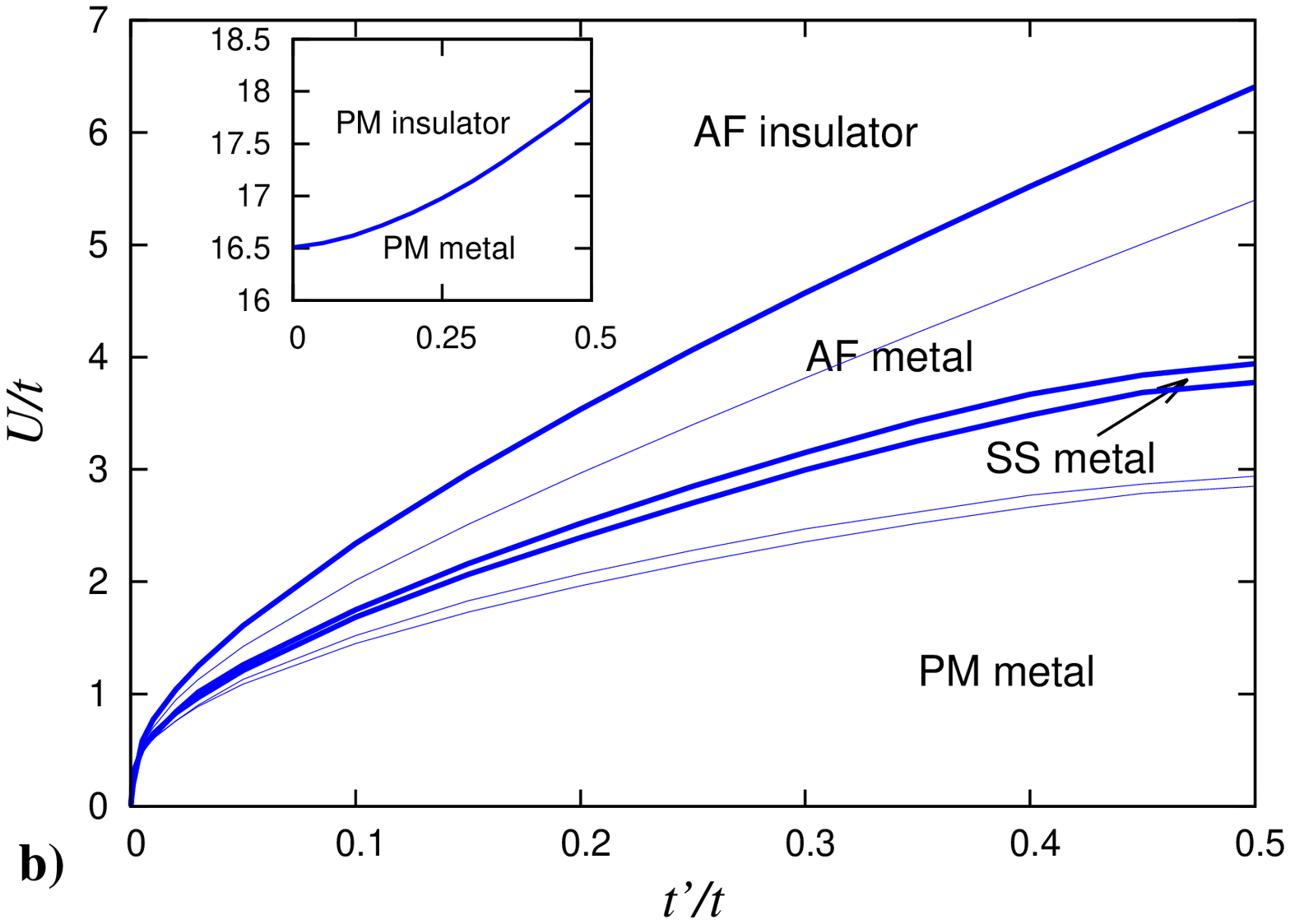}
  \caption{The same as in Fig. \ref{fig:square_phase_diagram} for a) simple cubic lattice, b) body-centered cubic lattice. Note that the phase labels refer only to SBA results.}
\label{fig:3d_phase_diagrams}
\end{figure}

We have calculated  MIT phase diagrams analogous to that of Sec. \ref{sec:square} for simple cubic (Fig.~\ref{fig:3d_phase_diagrams}a) and body-centered cubic (Fig.~\ref{fig:3d_phase_diagrams}b) lattices. 
In both these cases the metal--insulator transition is of second order and occurs in antiferromagnetic state when the gap in energy spectrum opens. 
The dramatic difference with the case of the square lattice is the presence of wide AF metal phase region. 
In the case of sc lattice we state (i) scenario for $t'\lesssim0.26t$, but when $t'\gtrsim0.26t$  this scenario is slightly modified by the inclusion of the very narrow spin-spiral metal phase region ($\mathbf{Q}=(\pi-\delta,\pi,\pi)$). The transition between spin-spiral state and AF metal is of the first order, and we denote it as (iv) scenario. 
For bcc lattice generally the (iv) scenario is realized ($\mathbf{Q}=(2\pi-\delta,2\pi-\delta,2\pi-\delta)$) with the only difference being the second order of the transition between the spin-spiral and AF metal phases.

One can see that the Brinkman--Rice transition takes place for much larger values of $U$ than the magnetic MIT, as well as for square lattice. In Ref. \cite{2004:Irkhin} a comparison of the paramagnetic MIT critical $U$'s obtained by the generalized Hubbard-III approximation and `linearized DMFT` is presented for $t'=0$. DMFT gives $U_c=12t$ for the square lattice, $U_c=14.6t$ for the simple cubic lattice and $U_c=17.0t$ for the bcc lattice which is close to our results (our BR  values of $U_c$ are $13.0t$, $16.1t$, $16.5t$, respectively). Hubbard--III values are noticeably smaller, but even they are much larger than the magnetic MIT transition points obtained in our research.

\section{Conlusions}
The phase diagrams of the Hubbard model at half-filling are constructed for the square, sc and bcc lattices using slave-boson and Hartree--Fock approximations. The role of spiral magnetic states strongly depends on the lattice type: 
for square and sc lattices the corresponding phase region appears at large $t'$, in the latter case it is very narrow; for bcc lattice the metallic spiral phase region is sizeable and appears at all $t'$. 
Three scenarios of the metal--insulator transition are found for the square lattice depending on the $t'$ value. The first one is the second-order antiferromagnetic metal--insulator transition, the second one is the first-order paramagnetic metal--antiferromagnetic insulator transition and the third one is the first-order spiral metal--antiferromagnetic insulator transition. In sc and bcc lattices MIT is a continuous transition from antiferromagnetic metal to antiferromagnetic insulator for all the $t'$ values studied.

This work was supported in part by the Division of Physical Sciences and Ural Branch, Russian Academy of Sciences
(project no. 15-8-2-9, 15-8-2-12) and by the Russian Foundation for Basic Research (project no.
14-02-31603-mol-a) and Act 211 Government of the Russian Federation 02.A03.21.0006.



%
%


\end{document}